\documentclass[preprint,preprintnumbers,amsmath,amssymb]{revtex4}
\pdfoutput=1
\usepackage{epsfig}
\usepackage{graphicx}
\usepackage{dcolumn}
\usepackage{bm}
\usepackage{threeparttable}
\usepackage{centernot}

\def\beq{\begin{equation}}
\def\eeq{\end{equation}}
\def\eeqn{\end{equation}}
\newcommand\iden{\leavevmode\hbox{\small1\normalsize\kern-.33em1}}

\newcommand{\bea} {\begin{eqnarray}}
\newcommand{\eea} {\end{eqnarray}}

\def\lam{\lambda}

\let\jnfont=\rm
\def\NPB#1 {{\jnfont Nucl.\ Phys.\ B }{\bf #1} }
\def\PLB#1 {{\jnfont Phys.\ Lett.\ B }{\bf #1} }
\def\EPJC#1 {{\jnfont Eur.\ Phys.\ Jour.\ C }{\bf #1} }
\def\PRD#1 {{\jnfont Phys.\ Rev.\ D }{\bf #1} }
\def\PRL#1 {{\jnfont Phys.\ Rev.\ Lett.\ }{\bf #1} }
\def\MPLA#1 {{\jnfont Mod.\ Phys.\ Lett.\ A }{\bf #1} }
\def\JPG#1 {{\jnfont J.\ Phys.\ G }{\bf #1} }
\def\CTP#1 {{\jnfont Commun.\ Theor.\ Phys.\ }{\bf #1} }
\def\JHEP#1 {{\jnfont JHEP \ }{\bf #1} }
\def\NPPS#1 {{\jnfont Nucl.\ Phys.\ Proc.\ Suppl.\ }{\bf #1} }
\def\CPC#1 {{\jnfont Comput.\ Phys.\ Commun.\ }{\bf #1} }
\def\CPL#1 {{\jnfont Chin.\ Phys.\ Lett. }{\bf #1} }
\def\APPB#1 {{\jnfont Acta\ Phys.\ Polon.\ B }{\bf #1} }

\def\lsim{\raise0.3ex\hbox{$<$\kern-0.75em\raise-1.1ex\hbox{$\sim$}}}
\def\gsim{\raise0.3ex\hbox{$>$\kern-0.75em\raise-1.1ex\hbox{$\sim$}}}
\def\PR#1 {{\jnfont Phys.\ Rept. }{\bf #1} }
\def\CHC#1 {{\jnfont Chin.\ Phys.\ C }{\bf #1} }
\def\NIMA#1 {{\jnfont Nucl.\ Instrum.\ Meth.\ A }{\bf #1} }
\def\JCAP#1 {{\jnfont JCAP \ }{\bf #1} }
\def\ASA#1 {{\jnfont Astron.\ Astrophys.\ A }{\bf #1} }  

\begin{document}

\title{\ \\[10mm] Dark matter, $Z^{\prime}$, vector-like quark at the LHC and $b\to s \mu\mu$ anomaly}

\author{Wei Chao$^{1}$, Hongxin Wang$^{2}$, Lei Wang$^{2}$, Yang Zhang$^{3}$}

\affiliation{$^1$ Center for Advanced Quantum Studies, Department of Physics, Beijing Normal University, Beijing, 100875, P. R. China \\
$^2$ Department of Physics, Yantai University, Yantai 264005, P. R. China\\
$^3$ School of Physics and Microelectronics, Zhengzhou University, ZhengZhou 450001, P. R. China}


\begin{abstract}
In this paper, combining the $b\to s\mu^+\mu^-$ anomaly and dark matter observables,
 we study the capability of LHC to test dark matter, $Z^{\prime}$, and vector-like quark. We focus on a local 
$U(1)_{L_\mu-L_\tau}$ model with a vector-like $SU(2)_L$ doublet quark $Q$ and a complex singlet scalar whose lightest component $X_I$ is
a candidate of dark matter. 
After imposing relevant constraints, we find that the $b\to s\mu^+\mu^-$ anomaly
and the relic abundance of dark matter favor $m_{X_I}< 350$ GeV and $m_{Z^{\prime}}< 450$ GeV for $m_Q<$ 2 TeV and $m_{X_R}<$ 2 TeV
(the heavy partner of $m_{X_I}$). The current searches for jets and missing transverse momentum 
at the LHC sizably reduce the mass ranges of the vector-like quark, and $m_Q$ is required to be larger than 1.7 TeV. 
Finally, we discuss the possibility
of probing these new particles at the high luminosity LHC via the QCD process $pp \to D\bar{D}$ or $pp \to U\bar{U}$ followed by the
decay $D\to  s (b) Z'X_I$ or $U \to u (c) Z' X_I$ and then $Z'\to\mu^+\mu^-$. Taking a benchmark point of $m_Q$=1.93 TeV, 
 $m_{Z^\prime}=170$ GeV, and $m_{X_I}=$ 145 GeV, we perform a detailed Monte Carlo simulation, and find that such benchmark point can be accessible at the 14 TeV LHC with an integrated luminosity 3000 fb$^{-1}$.

\end{abstract}

\maketitle

\section{introduction}
At present, there are several interesting excesses in $B$-physics
measurements involving the transition $b\to s\ell^+\ell^-$ ($\ell=\mu, e$),
\beq
R_{K^{(*)}} \equiv \frac{B \to K^{(*)}
\mu^+ \mu^-}{B \to K^{(*)} e^+ e^-}.
\eeq
The LHCb results for the $R_K$ ratio in one $q_2$ bin \cite{lhcb-rk-1,lhcb-rk-2} and the
$R_{K^*}$ ratio in two $q_2$ bins 
\cite{lhcb-rkstar} were found to lie significantly below one:
\begin{align}
R_K &= 0.846^{+0.060}_{-0.054}\text{(stat)}^{+0.016}_{-0.014}\text{(syst)}    \,, \quad
&q^2 \in [1,6]~\text{GeV}^2 \,, \nonumber \\[0.2cm]
R_{K^\ast} &= 0.660^{+0.110}_{-0.070}\text{(stat)}\pm0.024\text{(syst)}    \,, \quad
&q^2 \in [0.045,1.1]~\text{GeV}^2\,, \nonumber \\[0.2cm]
R_{K^\ast} &= 0.685^{+0.113}_{-0.069}\text{(stat)}\pm0.047\text{(syst)}     \,, \quad
&q^2 \in [1.1,6.0]~\text{GeV}^2 \,. 
\end{align}
Belle announced its measurement of $R_{K^*}$ \cite{belle-rkstar}
\beq
R_{K^*} =
\begin{cases}
0.52^{+0.36}_{-0.26} \pm 0.05 ~,~~ 0.045 \le q^2 \le 1.1 ~{\rm GeV}^2 ~, \\
0.96^{+0.45}_{-0.29} \pm 0.11 ~,~~ 1.1 \le q^2 \le 6.0 ~{\rm GeV}^2 ~, \\
0.90^{+0.27}_{-0.21} \pm 0.10 ~,~~ 0.1 \le q^2 \le 8.0 ~{\rm GeV}^2 ~, \\
1.18^{+0.52}_{-0.32} \pm 0.10 ~,~~ 15.0 \le q^2 \le 19.0 ~{\rm GeV}^2 ~, \\
0.94^{+0.17}_{-0.14} \pm 0.08 ~,~~ 0.045 \le q^2 ~.
\end{cases}
\eeq

The global fits to the experimental data show the new physics (NP) model can explain the anomalies of 
$R(K)$ and $R(K^{*})$ by contributing to $C_9^\mu$. With $C_{10}^{\mu,NP}$ =0, the best fit value
for $C_9^{\mu,NP}$ is $-1.10 \pm 0.16$ \cite{1903.10086}.


A $U(1)_{L_\mu-L_\tau}$ gauge boson couples only to $\mu(\tau)$ but
not to electron \cite{lu-lt}, and this type of $U(1)_{L_\mu-L_\tau}$ model has also been
modified from its minimal version to explain $b \to s\mu^+\mu^-$ anomaly \cite{bs1,bs2,bs3,bs4,bs5,bs6,bs7,bs8,bs9,bs10,bs11,bs12,bs13,bs14,bs15,bs16,1901.04761}.
In Ref. \cite{1901.04761}, in addition to the $U(1)_{L_\mu-L_\tau}$ gauge boson $Z'$ and 
a complex singlet ${\cal S}$ breaking $U(1)_{L_\mu-L_\tau}$ symmetry, 
a vector-like $SU(2)_L$ doublet quark $Q$ and a complex singlet $X$ are introduced to produce
the $Z'bs$ coupling large enough to explain the anomalies of $R(K^{(*)})$. As the lightest component of $X$, $X_I$ is
a candidate of dark matter (DM). In this paper, we will combine the $b\to s\mu^+\mu^-$ anomaly and the experimental data of DM,
 and study the capability of LHC to test dark matter, $Z^{\prime}$, and vector-like quark.

Our work is organized as follows. In Sec. II we recapitulate the
model. In Sec. III we consider the relevant theoretical constraints and $b\to s$ flavor observables,
 and explain the $b \to s\mu^+\mu^-$ anomaly. In Sec. IV, we discuss the DM observables.
In Sec. V, we use the current searches at the LHC to constrain the parameter space, and
analyze the possibility of probing the new particles at the high luminosity LHC.
Finally, we give our conclusion in Sec. VI.

\section{the model}
In addition to the $U(1)_{L_\mu-L_\tau}$ gauge boson $Z'$, the model predicts a complex singlet ${\cal S}$, a complex singlet $X$,
and a $SU(2)_L$ doublet quark $Q$. Their quantum numbers under the gauge group 
$SU(3)_C\times SU(2)_L\times U(1)_Y\times U(1)_{L_\mu-L_\tau}$
are shown in Table \ref{tabquantum}.

\begin{table}[t]
  \centering
  \caption{The quantum numbers of the vector-like quark $Q\equiv(U,D)$, the scalars $X$ and $S$ under the gauge group 
$SU(3)_C\times SU(2)_L\times U(1)_Y\times U(1)_{L_\mu-L_\tau}$.}
  \label{tabquantum}
  \begin{tabular}{ccccc}
  \hline\hline
            & SU(3)$_c$~~ & SU(2)$_L$ ~~& U(1)$_Y$ ~~& U(1)$_{B-L}$  \\ \hline
  $ Q $    & {\bf 3}   & {\bf 2}& $+1/6$ & $-q_x$  \\
  $ X $    & {\bf 1} & {\bf 1}& $0$ & $q_x$  \\
  $ {\cal S} $    & {\bf 1} & {\bf 1}& $0$ & $-2q_x$  \\ \hline
    \end{tabular}
\end{table}

The Lagrangian which remains invariant under the $SU(3)_C\times SU(2)_L\times U(1)_Y\times U(1)_{L_\mu-L_\tau}$ symmetry is given by 
\begin{align}
  {\cal L} &= {\cal L}_{\rm SM} -{1 \over 4} Z'_{\mu\nu} Z^{\prime\mu\nu}
              + g_{Z'} Z'^{\mu}(\bar{\mu}\gamma_\mu \mu + \bar{\nu}_{\mu_L}\gamma_\mu\nu_{\mu_L} - \bar{\tau}\gamma_\mu \tau - \bar{\nu}_{\tau_L}
             \gamma_\mu\nu_{\tau_L})\nonumber\\
            &-V  + \bar{Q} (i \centernot D -M_Q) Q\nonumber\\
            &+ (D_\mu X^\dagger) (D^\mu X)  + (D_\mu {\cal S}^\dagger) (D^\mu {\cal S}) 
  - \sum_{i=1}^3 (\lambda_i \bar{q}_L^i Q X + {h.c.}).
\label{eq:model}
\end{align} 
Where we ignore the kinetic mixing term of gauge bosons of $U(1)_{L_\mu-L_\tau}$ and $U(1)_Y$.
 $q^i_L$ denotes the SM left-handed quark doublet with $i=1,2,3$, and $D_\mu$ is the covariant derivative. 
The field strength tensor $Z'_{\mu\nu}=\partial_\mu Z'_\nu-\partial_\nu Z'_\mu$, and $g_{Z'}$ is the gauge coupling 
constant of the $U(1)_{L_\mu-L_\tau}$ group.
The scalar potential $V$ is given by
\begin{eqnarray} \label{scalarV} V &=& -\mu_{h}^2
(H^{\dagger} H) - \mu_{S}^2 ({\cal S}^{\dagger} {\cal S}) + m_X^2 (X^{\dagger} X) + \left[\mu X^2 {\cal S} + \rm h.c.\right]\nonumber \\
&&+ \lambda_H (H^{\dagger} H)^2 +
\lambda_S ({\cal S}^{\dagger} {\cal S})^2 + \lambda_X (X^{\dagger} X)^2 + \lambda_{SX}
({\cal S}^{\dagger} {\cal S})(X^{\dagger} X) \nonumber \\
&&+ \lambda_{HS}(H^{\dagger} H)({\cal S}^{\dagger} {\cal S}) + \lambda_{HX}(H^{\dagger} H)(X^{\dagger} X).
\end{eqnarray}
The SM Higgs doublet $H$, the singlet filed ${\cal S}$ and $X$ is expressed by
\begin{equation}
H=\left(\begin{array}{c} G^+ \\
\frac{1}{\sqrt{2}}\,(h_1+v_h+iG)
\end{array}\right)\,,
{\cal S}={1\over \sqrt{2}} \left( h_2+v_S+i\omega\right) \,,
X={1\over \sqrt{2}} \left( X_R+iX_I\right) \,,
\end{equation}
Where $v_h=246$ GeV and $v_S$ are respectively vacuum expectation values (VeVs) of $H$ and ${\cal S}$, 
and the $X$ field has no VeV.
The mass parameters $\mu^{2}_{h}$ and $\mu^{2}_{S}$ in the potential of Eq. (\ref{scalarV}) are determined by the potential minimization conditions,
\beq
\begin{split}
&\quad \mu_{h}^2 = \lam_H v_h^2 + {1 \over 2} \lam_{HS} v_S^2,\\
&\quad \mu_{S}^2 = \lam_S v_S^2 + {1 \over 2} \lam_{HS} v_h^2.\\
\end{split}
\label{min_cond}
\eeq
After ${\cal S}$ acquires the VeV, the $\mu$ term makes the complex scalar $X$ split into two real scalar fields $X_R$, $X_I$, and
their masses are given by
\begin{align}
 &m_{X_R}^2 = m_X^2 +  {1 \over 2} \lambda_{HX} v_H^2 + {1 \over 2}\lambda_{SX} v_S^2  + \sqrt{2} \mu v_S\nonumber\\
 &m_{X_I}^2 = m_X^2 +  {1 \over 2} \lambda_{HX} v_H^2 + {1 \over 2}\lambda_{SX} v_S^2  - \sqrt{2} \mu v_S.
\end{align}
The discrete $Z_2$ symmetry of the scalar potential in Eq. (\ref{scalarV}) makes the lightest component $X$
to be as a candidate of DM, which we assume is $X_I$.

The two physical CP-even states $h$ and $S$ are from the mixing of $h_1$ and $h_2$ by the following relation,
\begin{align} 
\left(
\begin{array}{c}
h_1 \\ h_2
\end{array}
\right)
=
\left(
\begin{array}{cc}
\cos\theta & \sin\theta\\
-\sin\theta & \cos\theta \\
\end{array}
\right)
\left(
\begin{array}{c}
h \\ S
\end{array}
\right),
\end{align} 
where $\theta$ is the mixing angle.
The two CP-even Higgses mediate the DM interactions,
\begin{align}
  {\cal L}(X_I X_I ,h,S)  =& -{1 \over 2} \Big[\lambda_{HX} v_H c_\theta- (\lambda_{SX} v_S-\sqrt{2} \mu) s_\theta \Big]  h X_I^2 \nonumber\\
                                        &-{1 \over 2} \Big[\lambda_{HX} v_H s_\theta  + (\lambda_{SX} v_S-\sqrt{2} \mu) c_\theta \Big]  S X_I^2.
\label{eq:HPcoupl}
\end{align}
In this paper, in order to suppress the stringent constraints from the DM direct detection and indirect detection 
experiments, we simply assume the $hX_IX_I$ coupling is absent, namely taking $\theta=0$ and $\lambda_{HX}=0$.
For $\theta=0$, we obtain the following expressions,
\beq
\lambda_{HS}=0,~~~\lambda_H=\frac{m_h^2}{2v_h^2},~~~\lambda_S=\frac{m_S^2}{2v_S^2}.
\eeq
After ${\cal S}$ gets VEV, the $U(1)_{L_\mu-L_\tau}$ gauge boson $Z'$ obtains a mass,
\beq m_Z' = 2g_Z' \mid q_x\mid v_S.
\eeq

The complex singlet $X$ mediates the new Yukawa interactions of the vector-like quarks and the SM left-handed quark,
\begin{align}
  \Delta {\cal L}_{\rm Yukawa} &= -{1 \over \sqrt{2}} \sum_{i=1,2,3} \left(\lambda_{u_i}  \bar{u}_{iL} U
                                 +\lambda_{d_i} \bar{d}_{Li} D \right)(X_R +i X_I)+h.c.,
\end{align}
where we assume that the down-type quarks are already in the mass basis, and rotate the interaction eigenstates of up-type quarks
to the mass eigenstates via the CKM matrix $V$. Thus, $\lambda_{u_i} \equiv \sum_j V_{ij} \lambda_j$ and $\lambda_{d_i} \equiv \lambda_i$
 with $u_i=u,c,t$ and $d_i=d,s,b$.
We will simply set $\lambda_1=0$ to remove the constraints related to the first generation quarks. 
 As a result, $\lambda_u$ is much smaller than $\lambda_c$ and $\lambda_t$ due to the suppression of
the factors of $V_{us}$ and $V_{ub}$.

\section{$b\to s \mu^+\mu^-$ anomaly}
 We apply the upper bound of $g_{Z'}/m_{Z'}\leq$ (550 GeV)$^{-1}$ from the neutrino trident process \cite{trident}, 
and require $g_{Z'}q_x \leq 1$ to maintain the perturbativity of the $Z'$ couplings. 
The tree-level stability of the potential of Eq. (\ref{scalarV}) requires
\begin{eqnarray}
&&\lambda_H \geq 0 \,, \quad \lambda_S \geq 0 \,,\quad \lambda_X \geq 0 \,,\quad \nonumber \\
&& \lambda_{HS} \geq - 2\sqrt{\lambda_H \,\lambda_{S}}  \,, \quad
\lambda_{HX} \geq - 2\sqrt{\lambda_H \,\lambda_{X}}  \,, \quad
\lambda_{SX} \geq - 2\sqrt{\lambda_S \,\lambda_{X}} \,, \quad\nonumber \\
&&\sqrt{\lambda_{HS}+2\sqrt{\lambda_H \,\lambda_S}}~\sqrt{\lambda_{HX}+
2\sqrt{\lambda_H \,\lambda_{X}}}
~\sqrt{\lambda_{SX}+2\sqrt{\lambda_S\,\lambda_{X}}} \nonumber \\
&&+ 2\,\sqrt{\lambda_H \lambda_S \lambda_{X}} + \lambda_{HS} \sqrt{\lambda_{X}}
+ \lambda_{HX} \sqrt{\lambda_S} + \lambda_{SX} \sqrt{\lambda_H} \geq 0 \,.
\end{eqnarray}

We scan over the other parameters in the following ranges:
\begin{align}
 & 60 {\rm GeV} \leq m_{X_I} \leq 1 {\rm TeV},~~~ 800 {\rm GeV} \leq m_{X_R} \leq 2 {\rm TeV},~~~1 {\rm TeV} \leq m_Q \leq 2 {\rm TeV},\nonumber\\
 & 100 {\rm GeV} \leq m_{Z'} \leq 1000 {\rm GeV},~~~100 {\rm GeV} \leq m_{S} \leq 1000 {\rm GeV},\nonumber\\
 & 0.1 \leq \lambda_{bs}(\equiv\lambda_b \lambda_s) < 0.3~{\rm with}~ \lambda_{b,s} \leq 1. 
\end{align}

We consider four relevant $b\to s$ flavor observables, $R_{K^{(*)}}$, $\Delta m_s$, $B\to X_s\gamma$, and $R_{K^{(*)}}^{\nu\nu}$, which are
introduced in detail in Ref. \cite{1901.04761}. Here we give the expressions for calculating the four observables briefly.

\subsection{Numerical calculations}
{\bf I. $R_{K^{(*)}}$ anomalies}

The model does not contain the tree-level $Z'$-$b$-$s$ flavor-changing coupling, but produces the $Z'$-$b$-$s$ coupling via
the one-loop involving the vector-like quarks, $X_R$ and $X_I$. The $b\to s \mu^+\mu^-$ transition operator $O_9^\mu$ is 
generated by $Z'$-exchanging penguin diagrams. The corresponding Wilson coefficient $C_9^{\mu,NP}$ is given by \cite{1901.04761},
 \begin{align}
   C_9^{\mu,{\rm NP}} = -\frac{\sqrt{2} q_x}{8 G_F m_{Z'}^2} \, \frac{\alpha_{Z'}}{\alpha_{\rm em}} \,\frac{\lambda_s \lambda_b^*}{ V_{ts}^*
   V_{tb} }   \Bigg[ {1 \over 2}(k'(x_I)+k'(x_R))-k(x_I,x_R)\Bigg],
\label{eq:C9NP}
 \end{align}
where $x_{R,I}=m^2_{X_{R,I}}/m^2_{Q}$,
\begin{align}
k(x) &= \frac{ x^2 \log x}{x-1},~~~k(x_1,x_2) = \frac{k(x_1) - k(x_2)}{x_1 - x_2}. 
\end{align}
The prime on the $k$ functions denotes a derivative with respect to the argument.
A large mass splitting between $m_{X_R}$ and $m_{X_I}$ can enhance the absolute value of $C_9^{\mu,NP}$ which can explain 
$R_{K^{(*)}}$ anomaly.

{\bf II. $\Delta m_s$ for $B_s-\bar{B}_s$ mixing, $B\to X_s\gamma$, and $R_{K^{(*)}}^{\nu\nu}$}

The model gives the new contributions to $B_s-\bar{B}_s$ mixing via the box diagrams involving the vector-like quarks, $X_R$ and $X_I$, 
which can be written in the form
\beq
H_{eff}^{\Delta B=2, NP}= C_1^{NP}  (\bar{s} \gamma_\mu P_L b) (\bar{s}  \gamma^\mu P_L b). 
\eeq
Where $C_1^{NP}$ is given as \cite{1901.04761}
\begin{align}
C_1^{\rm NP} &= \frac{(\lambda_s \lambda_b^*)^2}{128 \pi^2 M_D^2} k(1,x_R,x_I),
\label{eq:mbs}
\end{align}
where 
\beq
k(1,x_R,x_I)=\frac{k(1,x_I) - k(x_R,x_I)}{1 - x_R}.
\eeq

At the $2\sigma$ confidence level, the measurement of the mass difference in
the $B_s-\bar{B}_s$ system gives a constraint on the value of $C_1^{NP}$ \cite{bs7},
\begin{align}
-2.1 \times 10^{-11} \le C_1^{\rm NP} \le 0.6 \times 10^{-11} \, ({\rm GeV}^{-2}).
\label{eq:BsBs}
\end{align}

The model gives the new contributions to $B\to X_s\gamma$ via the one-loop diagram involving 
the vector-like quarks, $X_R$ and $X_I$. The Wilson coefficients $C_{7\gamma,8g}$ is corrected \cite{1901.04761},
\begin{align}
C_{7\gamma}^{\rm NP}&= \frac{\sqrt{2}}{48} \frac{\lambda_s \lambda_b^*}{V_{ts}^* V_{tb}} \frac{1}{G_F M_D^2}
                      \left(J_1(x_I)+J_1(x_R)\right), \nonumber\\
C_{8g}^{\rm NP}&=  -\frac{\sqrt{2}}{16} \frac{\lambda_s \lambda_b^*}{V_{ts}^* V_{tb}} \frac{1}{G_F M_D^2}
                      \left(J_1(x_I)+J_1(x_R)\right), 
\end{align}
where 
\begin{align}
J_1(x) &=\frac{1-6x+3x^2+2x^3-6x^2 \log x}{12(1-x)^4}.
\end{align}
The experimental measurement of the inclusive branching fraction of $B\to X_s\gamma$ is $(3.32\pm 0.15)\times 10^{-4}$ \cite{bsr-exp},
 and the SM prediction is $(3.36\pm 0.23) \times 10^{-4}$ \cite{bsr-sm}. The explanation of experimental values at $2\sigma$ level requires
\begin{align}
-6.3 \times 10^{-2} \le C_{7\gamma}^{\rm NP}+0.24\, C_{8g}^{\rm NP} \le 7.3 \times 10^{-2}.
\label{eq:bsr_constraint}
\end{align}

The model gives the additional contributions to $B\to K^{(*)} \nu\bar{\nu}$ via the diagrams which 
are obtained by replacing the external muon lines of the $b\to s \mu^+\mu^-$ diagrams with the neutrino lines. 
The current experimental bounds are
\begin{align}
R_K^{\nu\bar{\nu}} < 4.3, \quad R_{K^*}^{\nu\bar{\nu}} < 4.4, \quad (\text{at 90\% C.L.}).
\end{align}
with \begin{align}
R_{K^{(*)}}^{\nu\bar{\nu}} =\frac{{\cal B}(B \to K^{(*)} \nu \bar{\nu})^{\rm exp}}{{\cal B}(B \to K^{(*)} \nu \bar{\nu})^{\rm SM}}.
\end{align}
In the model, the prediction value of $R_{K^{(*)}}^{\nu\nu}$ is \cite{1901.04761}
\begin{align} 
R_{K^{(*)}}^{\nu\bar{\nu}} &=\frac{\sum_{i=1}^3\left|C_L^{\rm SM} + C_L^{ii,{\rm NP}}\right|^2}{3 \left|C_L^{\rm SM} \right|^2}
=1+  \frac{2 \left|C_L^{22,{\rm NP}} \right|^2}{3 \left|C_L^{\rm SM} \right|^2},
\end{align}
with $C_L^{\rm SM} \approx -6.35$, $C_L^{11,NP}=0$, and
 \begin{align}
   C_L^{22,NP}=-C_L^{33,NP}= -\frac{\sqrt{2} q_x}{16 G_F m_{Z'}^2} \, \frac{\alpha_{Z'}}{\alpha_{\rm em}} \,
\frac{\lambda_s \lambda_b^*}{ V_{ts}^*   V_{tb} }   \Bigg[ {1 \over 2}(k'(x_I)+k'(x_R))-k(x_I,x_R)\Bigg].
\label{eq:CnuNP}
 \end{align}

\subsection{Results and discussions}
\begin{figure}
\vspace{-3.0cm}
 \epsfig{file=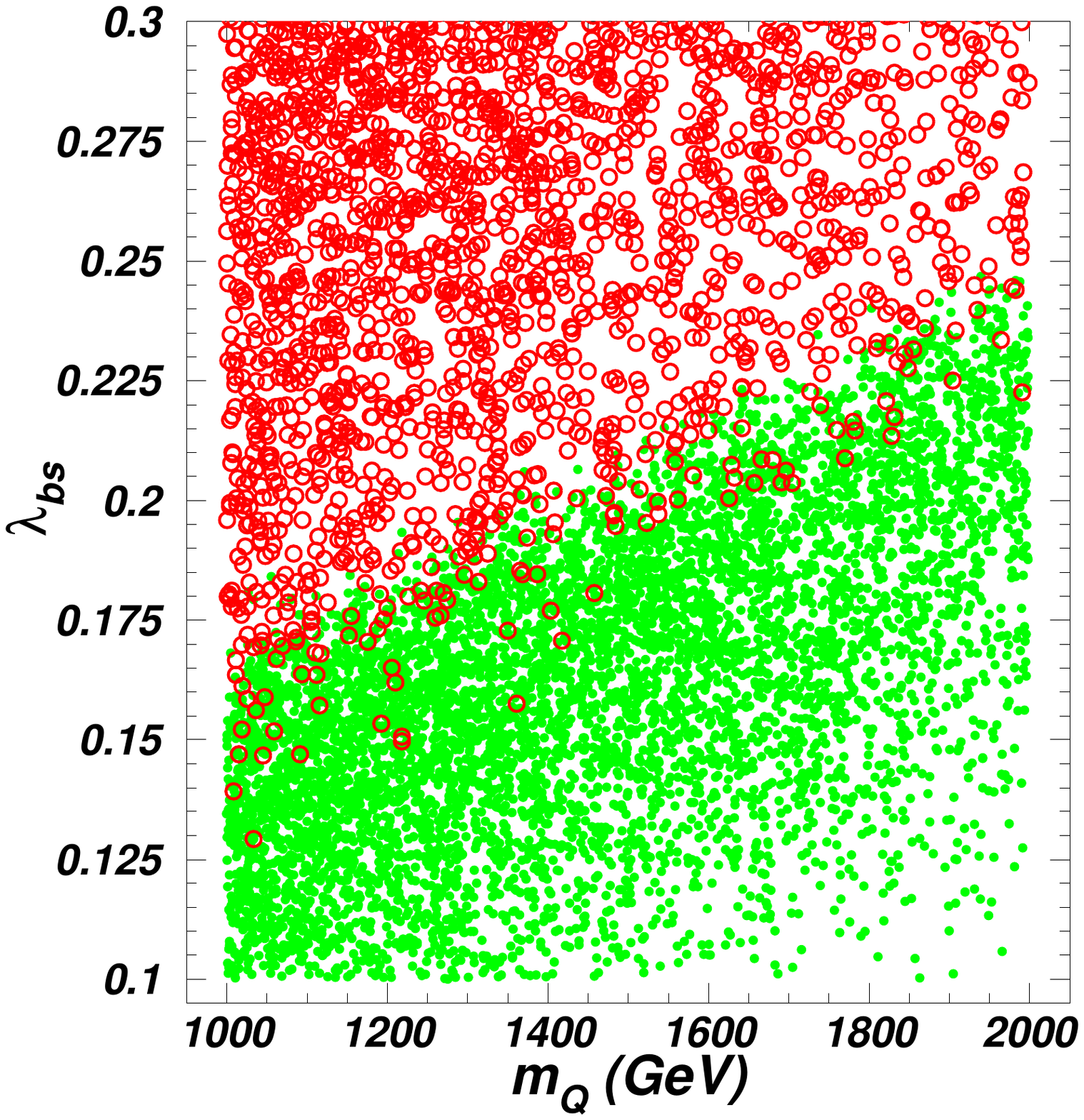,height=12.cm}
\vspace{-2.5cm} \caption{The surviving samples projected on the
planes of $m_Q$ versus $\lambda_{bs}$. All the samples accomodate the $R_{K^{(*)}}$ anomaly, and the bullets (green) and 
circles (red) are respectively allowed and excluded by the $\Delta m_s$.} \label{figrk}
\end{figure}

After imposing the constraints mentioned above, we use the model to explain the $R_{K^{(*)}}$ anomalies.
The bounds of $B\to X_s\gamma$ and $R_{K^{(*)}}^{\nu\nu}$ are almost satisfied in  the whole parameter space being consistent with $R_{K^{(*)}}$.
However, there is a strong correlation between $\Delta m_s$ and $R_{K^{(*)}}$, as shown in the Eq. (\ref{eq:C9NP}) and Eq. (\ref{eq:mbs}). 
Fig. \ref{figrk} shows that $R_{K^{(*)}}$ are explained in the whole region of 1000 GeV $\leq m_Q\leq$ 2000 GeV and 0.1 $\leq\lambda_{bs}\leq$ 0.3.
However, $\Delta m_s$ imposes an upper bound on $\lambda_{bs}$, which increases with $m_Q$. 
Due to the constraints of $\Delta m_s$, the $R_{K^{(*)}}$ anomaly can be only explained in the region of $\lambda_{bs}\leq$ 0.25.

\begin{figure}
 \epsfig{file=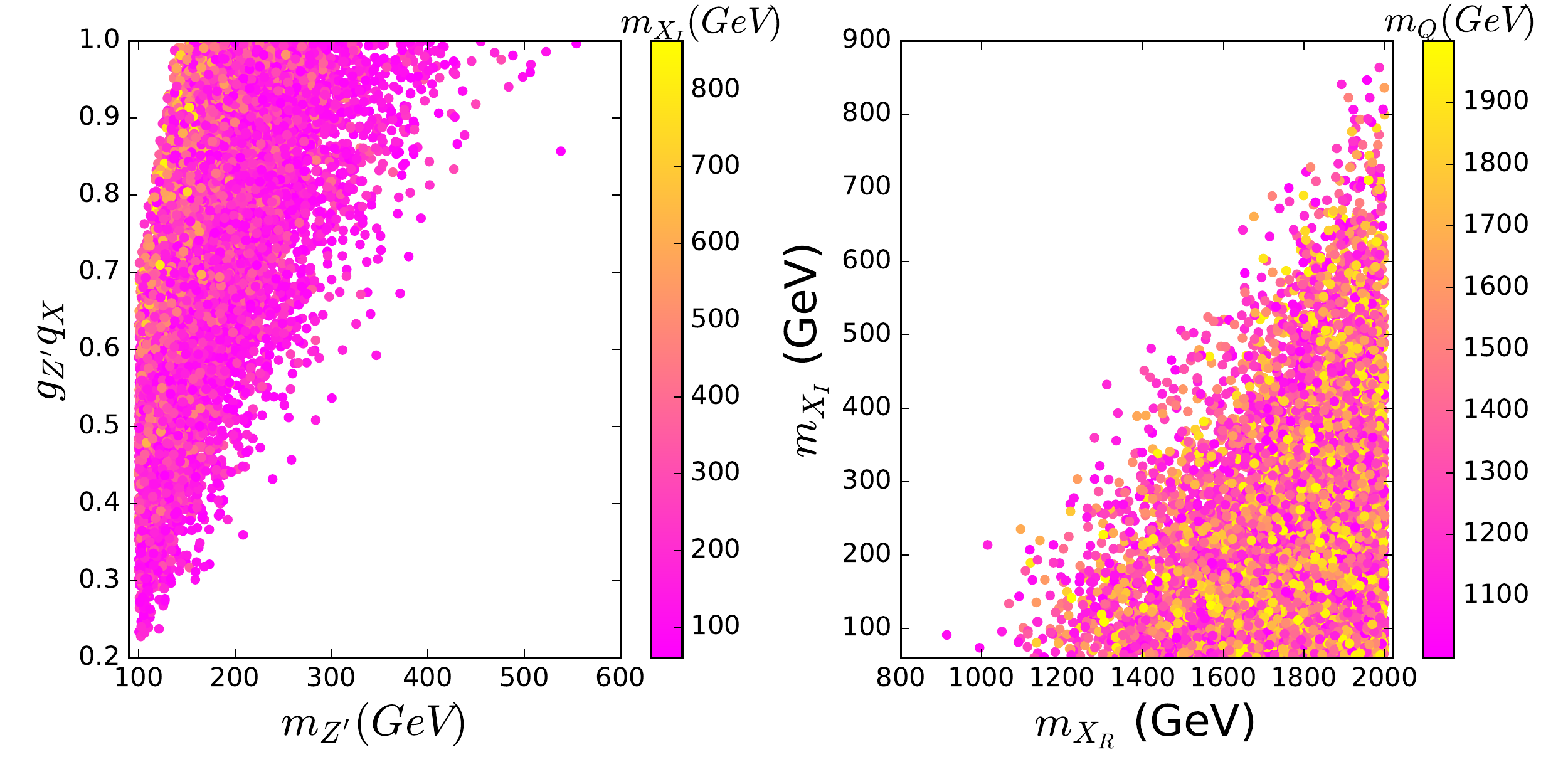,height=8.cm}
\vspace{-0.5cm} \caption{All the samples accomodate the $R_{K^{(*)}}$ anomaly, and satisfy the relevant $b\to s$ flavor observables, the neutrino trident process,
and the theoretical constraints.} \label{figrk2}
\end{figure}
After imposing the relevant $b\to s$ flavor observables, the neutrino trident process,
and the theoretical constraints, the samples explaining the $R_{K^{(*)}}$ anomaly are projected on the Fig. \ref{figrk2}. 
The left panel shows that the parameters $g_{Z'}q_X$ and $m_{Z'}$ are imposed strong constraints. Due to the constraints of the neutrino trident process,
the region with small $m_{Z'}$ and large $g_{Z'}q_X$ is empty. To accomodate the $R_{K^{(*)}}$ anomaly, $m_{Z'}$ is required to increase with $g_{Z'}q_X$.
Since we take $g_{Z'}q_x \leq 1$ to maintain the perturbativity of the $Z'$ couplings, $m_{Z'} > $ 600 GeV is excluded. 
Similarly, $g_{Z'}q_x \leq 0.2$ is disfavored since the minimal value of $m_{Z'}$ is taken as 100 GeV.

The right panel of Fig. \ref{figrk2} shows that $m_{X_I}$ is required to increase with $m_{X_R}$ since a sizable mass splitting between $m_{X_R}$ and $m_{X_I}$ is favoured to explain the $R_{K^{(*)}}$ anomaly. Because we choose $m_{X_R}\leq$ 2 TeV, $m_{X_I}$ is
required to be smaller than 900 GeV. Similarly, $m_{X_R}\leq$ 800 GeV is disfavored since the minimal value of $m_{X_I}$ is taken as 60 GeV. 

\section{Dark matter}
\begin{figure}
\hspace*{-1.2cm}
 \epsfig{file=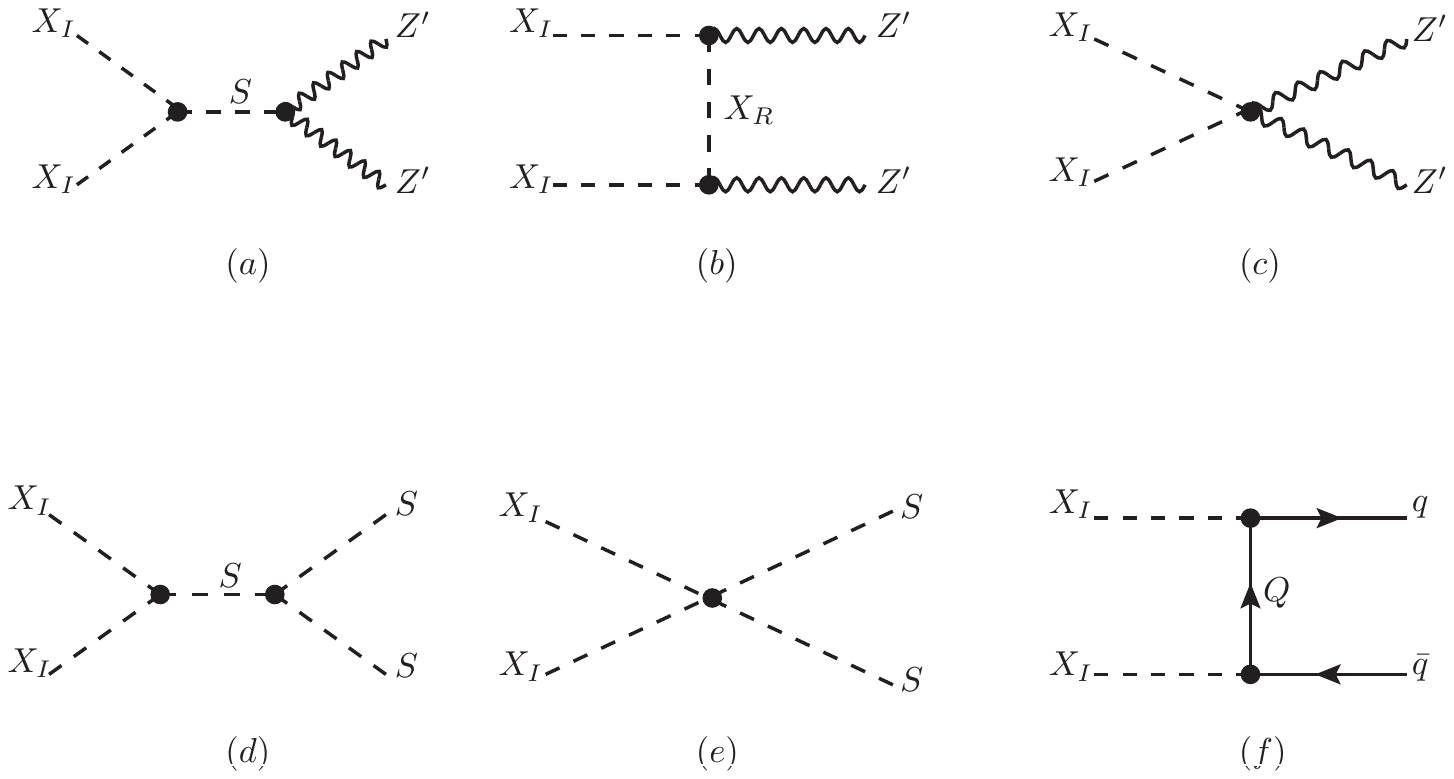,height=28.cm}
\vspace{-19.0cm} \caption{The Feynman diagrams for $X_IX_I\to Z'Z',~SS,~q\bar{q}$.} \label{figfmt}
\end{figure}

In the chosen parameter space, the DM can annihilation into $Z'Z'$, $SS$, and the SM quarks.
The corresponding Feynman diagrams are shown in the Fig. \ref{figfmt}.
The $X_IX_I\to q\bar{q}$ processes proceed through the $D(U)$-exchanging t-channel diagrams.
For 1 TeV $\leq m_Q\leq$ 2 TeV, $\lambda_b<1$, and $\lambda_s<1$, the annihilation cross sections are very small,
and their contributions to the relic density can be ignored. The $X_IX_I\to SS$ processes proceed through 
the $S$-exchanging s-channel diagram and the diagram of the quartic coupling $X_IX_ISS$.
The $X_IX_I\to Z'Z'$ proceed through the $S$-exchanging s-channel diagram, the $X_R$-exchanging t-channel diagram,
 and the diagram of the quartic coupling $X_IX_I Z'Z'$. 

We use $\textsf{micrOMEGAs}$ \cite{micomega} to calculate the relic density and the spin-independent DM-nucleon cross section.
The model file is generated by $\textsf{FeynRules}$ \cite{feyrule}. The Planck collaboration reported the relic density of 
cold DM in the universe,
 $\Omega_{c}h^2 = 0.1198 \pm 0.0015$ \cite{planck}.

The annihilation cross section of $X_IX_I\to Z'Z'$ from the diagram of Fig. \ref{figfmt}(c) only depends on 
three parameters $g_{Z'}q_X$,  $m_{Z'}$, $m_{X_I}$.
Since the $R_{K^{(*)}}$ anomaly imposes a lower bound on $g_{Z'}q_X$, for $m_{Z'}<m_{X_I}$ the annihilation cross sections of $X_IX_I\to Z'Z'$ 
are much larger than the value producing the correct relic density. Similarly, for $m_{S}<m_{X_I}$ 
the annihilation cross sections of $X_IX_I\to SS$ 
are too large to obtain the correct relic density. Therefore, we need to use the effects of forbidden channel 
to produce the relic density, namely that 
$m_{Z'}$ or $m_{S}$ is appropriately larger than $m_{X_I}$. In the calculation of the thermal averaged cross section,
the kinetic energy of the DM is nonnegligible in the early universe. When the mass difference is not too large and the DMs move fast,
the center of mass energy exceeds twice $m_{Z'}$ or $m_{S}$. Therefore, the process $X_IX_I\to Z'Z'~(SS)$ can occur in the early universe when 
$m_{X_I}$ has appropriate mass difference from $m_{Z'}$ ($m_S$). In addition, the temperature at the present time is much lower than the freeze-out temperature, and the velocity of DM is much smaller than that in the early universe. The channel $X_IX_I\to Z'Z'~(SS)$ are kinematically forbidden at the present time,
therefore the experimental constraints of the indirect detection of DM can be naturally satisfied. 

\begin{figure}
 \epsfig{file=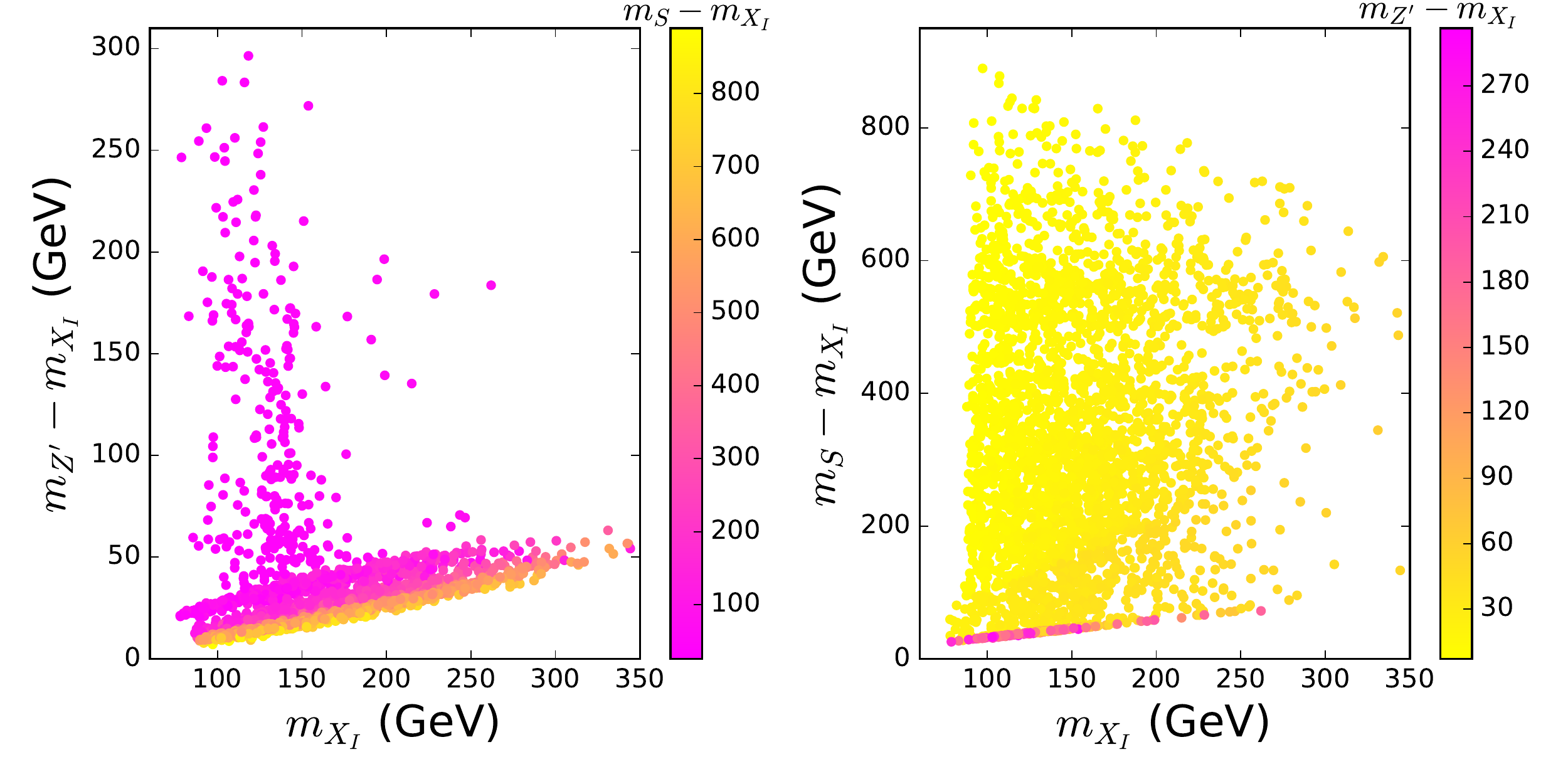,height=8.cm}
\vspace{-0.5cm} \caption{The surviving samples satisfying the DM relic density and the constraints of "pre-DM".} \label{figdm}
\end{figure}

After imposing the constraints of "pre-DM" (denoting the $R_{K^{(*)}}$ anomaly, the relevant $b\to s$ flavor observables, the neutrino trident process,
and the theoretical constraints), we find some samples which can achieve the correct DM relic density. The surviving samples are project on the Fig. \ref{figdm}. 
From the left panel, we find that the relic density favors $m_{X_I}<$ 350 GeV, and most of the surviving
 samples lie in the region of $m_{Z'}-m_{X_I}<60$ GeV.
For a large $m_S$, the annihilation cross section of $X_IX_I\to Z'Z'$ from the diagram of Fig. \ref{figfmt}(a)
 is suppressed. Therefore, a small value of $m_{Z'}-m_{X_I}$ is required to
enhance the cross section. For a large value of $m_{Z'}-m_{X_I}$, the $X_IX_I\to Z'Z'$ channel is still forbidden in the early universe, and does not
contribute to the relic density. For such case, the $X_IX_I\to SS$ channel will play the dominant contribution to the relic density. As shown in
the right panel, for a large value of $m_{Z'}-m_{X_I}$, a small value of $m_{S}-m_{X_I}$ is required to open the $X_IX_I\to SS$ channel in the 
early universe.

Exchanging an initial state $X_I$ and a final state quark of Fig. \ref{figfmt}(f),
we can obtain the Feynman diagrams which give the contributions to the cross section of the DM scattering off the nuclei. 
In the chosen parameter space, we find that the bounds of the XENON1T fail to exclude the parameter space 
achieving the correct relic density \cite{xenon2018}.

\section{the dark matter, $Z'$, and vector-like quark at the LHC}
\subsection{The current constraints from the direct searches at the LHC}
At the LHC, the vector-like quarks $D$ and $U$ are produced in pairs via the QCD processes, 
\beq
pp\to D\bar{D}, U\bar{U}.
\eeq
In the chosen parameter space, the $D$ and $U$ have following decay modes,
\beq
D\to X_I d_i, X_R d_i,~~~U\to X_I u_i, X_R u_i
\eeq
with 
\beq
X_R\to X_I Z'\to X_I \mu^+ \mu^-, X_I\tau^+\tau^-, X_I\nu_\mu \bar{\nu}_\mu,  X_I\nu_\tau \bar{\nu}_\tau.
\eeq
Since the $R_{K^{(*)}}$ anomaly and the DM relic density favor 
$X_R$ to be much larger than $X_I$, $D$ and $U$ will mainly decay into
$X_Is$, $X_Ib$, and $X_Iu_i$. In this paper, the coupling of $X_I$ and $d$ quark is taken as zero.

\begin{figure}
 \epsfig{file=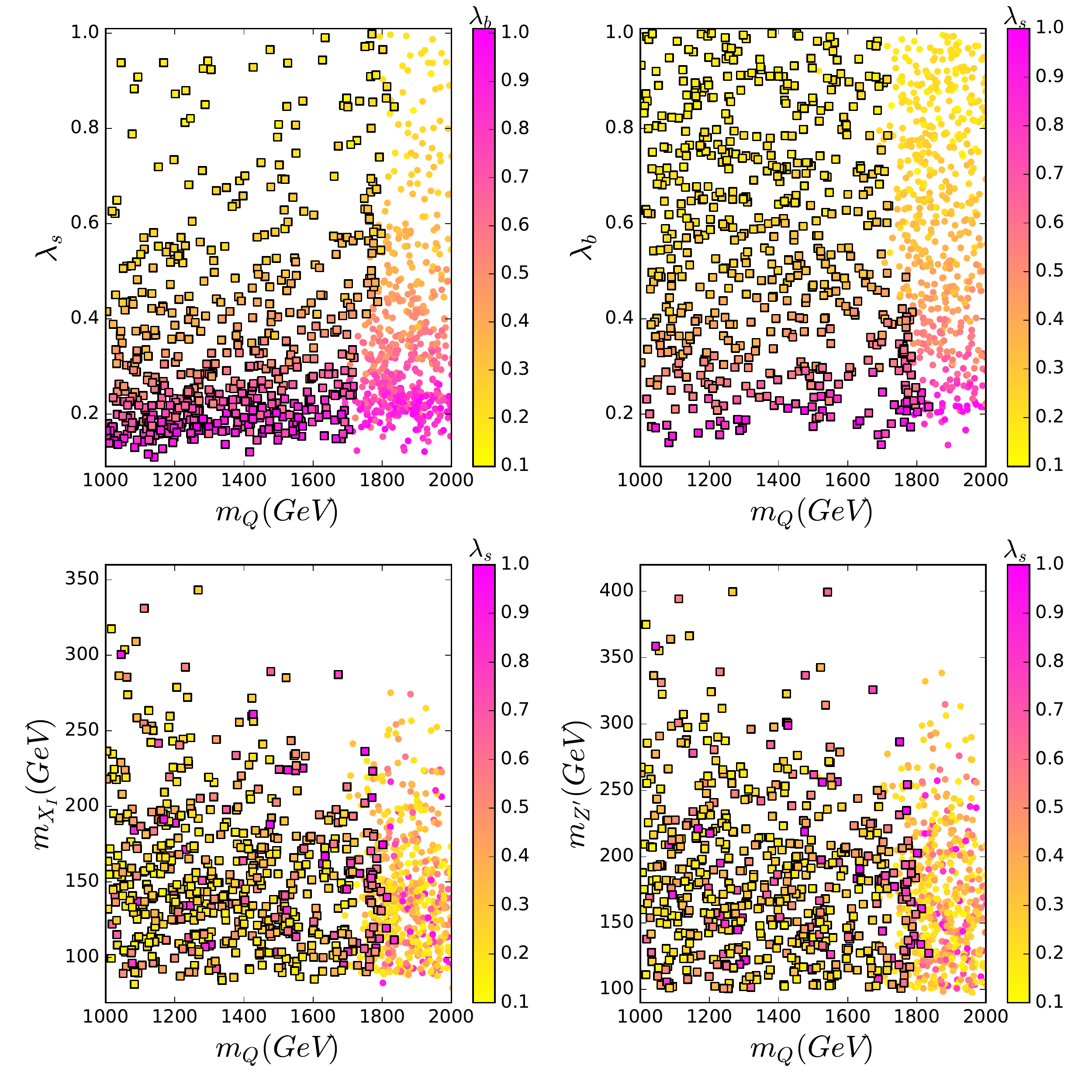,height=14.cm}
\vspace{-0.5cm} \caption{All the samples satisfy the constraints of "pre-DM" and the DM observables.
The squares and bullets are respectively excluded and allowed by the current direct searches at the LHC.} \label{figlhc}
\end{figure}
In order to restrict the productions of the above processes at the LHC for our model, 
we perform simulations for the samples 
using \texttt{MG5\_aMC-2.7.3}~\cite{Alwall:2014hca} 
with \texttt{PYTHIA8}~\cite{Torrielli:2010aw} and 
\texttt{Delphes-3.2.0}~\cite{deFavereau:2013fsa}, and adopt the constraints 
from all the analysis for the 13 TeV LHC in 
version \texttt{CheckMATE 2.0.28}~\cite{Dercks:2016npn}. 
For the excluded samples, the most sensitive experimental analysis is the ATLAS search for the squarks and gluinos in
final states containing jets and missing transverse momentum at 13 TeV LHC with 139 fb$^{-1}$ integrated luminosity data ~\cite{atlas-2019-040}. 
The final states $E_T^{miss}$ + $jets$ are just the main signal of the $D\bar{D}$ and $U\bar{U}$ in the model.

In Fig. \ref{figlhc}, all the samples satisfy the constraints of "pre-DM" and the DM observables. The current direct searches at the LHC
exclude $m_Q<$ 1.7 TeV. For a large $\lambda_s$, some samples with $m_Q$ around 1.8 TeV can be also excluded.
With an increase of $m_Q$, the production cross sections of $pp\to D\bar{D},~U\bar{U}$ are suppressed by the phase space,
and the direct searches at the LHC can be satisfied.

\subsection{The searches for the new particles at the high luminosity LHC}
Since the vector-like quark $U$ and $D$ are charged under the $U(1)_{L_\mu-L_\tau}$, 
the gauge boson $Z'$ has the tree-level couplings to the vector-like quarks.
Therefore, the model provides a novel approach of searching for $Z'$, the vector-like quark, and DM. $Z'$ is 
produced via the QCD process $pp \to D\bar{D}$ or $pp \to U\bar{U}$ followed by the
decay $D\to  s (b) X_R\to s(b)Z'X_I$ or $U \to u (c) X_R\to u (c) Z' X_I$, and then decays into $\mu^+\mu^-$. 

We pick a benchmark point which accomodates the $b\to s\mu^+\mu^-$ anomaly, and satisfies the constraints of "pre-DM", 
the DM observables, and the current
searches at the LHC. Several key input and output parameters are shown in Table \ref{tablhc}.


\begin{table}
\begin{footnotesize}
\begin{tabular}{| c | c | c | c | c| c| c| c| c| c|}
\hline
$m_{Z'}$(GeV) &$m_{X_I}$(GeV) &$m_{X_R}$(GeV) &$m_Q$(GeV) & $Br(D\to X_Ib)$ &$Br(D\to X_Is)$&$Br(D\to X_Rb)$
&$Br(D\to X_Rs)$\\
\hline
170 & 145  & 1309 & 1930 & 0.63  & 0.14 & 0.19 &0.04 \\
\hline
\end{tabular}
\end{footnotesize}
\caption{Several key input and output parameters for the benchmark point.}
\label{tablhc}
\end{table}


Now we perform detailed simulations on the signal and backgrounds at the 14 TeV LHC with high luminosity. 
 We choose the signal to
contain opposite sign di-muon ($\mu^+\mu^-$), missing transverse momentum $E^{miss}_T$, and multijet ($\geq$ 2 jets)
which include at least one $b$-jet.
The major SM irreducible background processes to this signal are $t\bar{t}$, $WW+$ jets, $ZZ+$ jets, and $WZ+$ jets.

We identify the muon candidates by requiring them to have $p_T>15$
GeV and $|\eta|<2.5$. The anti-kt algorithm is employed to
reconstruct the jets with a radius parameter $R = 0.4$ \cite{kt},
and the jets are required to have $p_T>20$ GeV and $|\eta|<2.5$. We
assume an average $b$-tagging efficiency of 80\% for real $b$-jets.

In order to suppress the contributions from the SM process,
we apply the "stransverse" mass, $m_{T2}$ \cite{mt2-1,mt2-2,08105178}, defined as
\beq
m_{T2}=\min_{\bf q_T}\left[\max\left(m_T({\bf p_T^{\ell_1},q_T}),m_T({\bf p_T^{\ell_2},p_T^{miss}-q_T})\right)\right]
\eeq  
where ${\bf p_T^{\ell_1}}$ and ${\bf p_T^{\ell_2}}$  are the transverse momenta of the di-muon. ${\bf q_T}$ is a transverse vector that
minimizes the larger of the two transverse masses $m_T$,
\beq
m_T({\bf p_T,q_T})=\sqrt{2(p_Tq_T-{\bf p_T\cdot q_T})}.
\eeq

\begin{figure}
 \epsfig{file=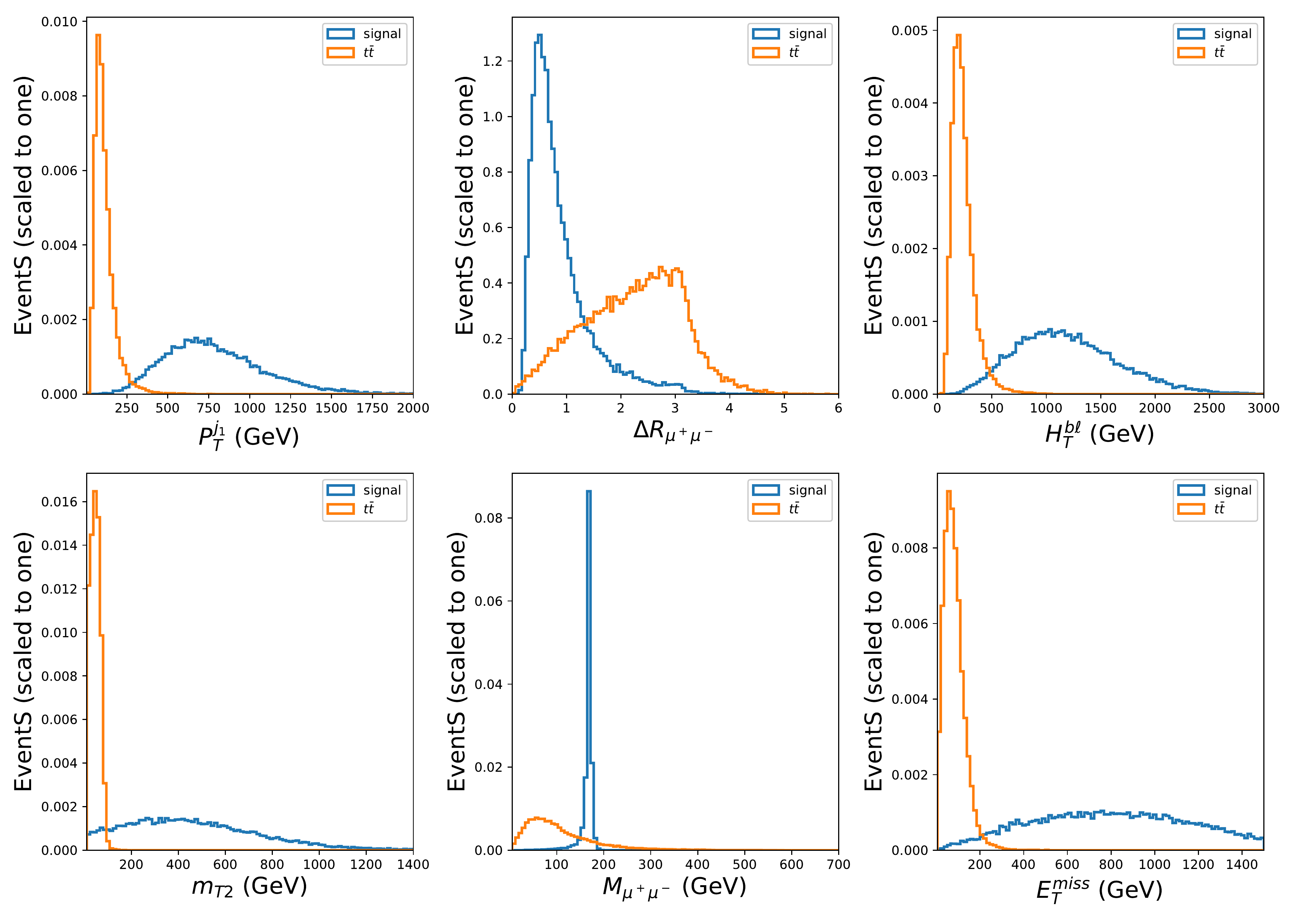,height=10.cm}
\vspace{-0.5cm} \caption{The signal and the $t\bar{t}$
background distributions of $P_T^{j_1,}$, $\Delta
R_{\mu^+\mu^-}$, $H_T^{b\ell}$, $m_{T2}$, $M_{\mu^+\mu^-}$, and $E_T^{miss}$ at the
14 TeV LHC, after requiring an opposite sign di-muon and multijet ($\geq$ 2 jets)
which include at least one $b$-jet.} \label{fig-dist}
\end{figure}

Fig. \ref{fig-dist} shows the distributions of some kinematical
variables at the LHC with $\sqrt{s}=14$ TeV for the signal and the background $t\bar{t}$.
The other processes are not shown since they are subdominant. 
According to the distribution
differences between the signal and backgrounds, we can improve the
ratio of signal to backgrounds by making some kinematical cuts.
We impose the following cuts
\bea &&P_T^{j_1}
> 290~{\rm GeV},~~P_T^{j_2} > 60~{\rm GeV},~~P_T^{b_1} >  60~{\rm GeV},\nonumber\\
&&\Delta R_{\mu^+\mu^-} < 2.0,~~~150 {\rm GeV} < M_{\mu^+\mu^-} < 180 {\rm GeV},\nonumber\\
&&E_T^{miss} > 310 {\rm GeV},~~~m_{T2} > 100 {\rm GeV},~~~H_T^{b\ell} > 500 {\rm GeV}.\label{basiccut}
\eea
Where $P_T^{j_1}$ and $P_T^{j_2}$ denote the transverse momentum of
the hardest and the second hardest jets which include $b$-jet, and $P_T^{b_1}$ denotes the transverse momentum 
of the hardest $b$-jet.
 $\Delta R = \sqrt{(\Delta
\phi)^2+(\Delta \eta)^2}$ is the particle separation with $\Delta
\phi$ and $\Delta \eta$ being the separation in the azimuthal angle
and rapidity respectively. $M_{\mu^+\mu^-}$ is the invariant mass of $\mu^+$ and $\mu^-$, and $H_T^{b\ell}$ is scalar sum of
transverse momentums of all the $b$-jets, $\mu^\pm$. Since $\mu^+$ and $\mu^-$ of the signal are 
from the decay of $Z^\prime$ with a mass of $170$ GeV, $M_{\mu^+\mu^-}$ appears a peak at 170 GeV, 
and $\Delta R_{\mu^+\mu^-}$ favors a small value.
The jets, $X_I$ and $\mu^\pm$ of the signal are the decay products of the vector-like quark with a mass of 1930 GeV, and such heavy mass
leads that these products tend to have large transverse momentums. The distributions of $m_{T2}$ for $t\bar{t}$ and 
$WW$+jets backgrounds peak before $m_W$. In addition, the DM $X_I$ has a mass of 145 GeV, 
therefore the signal events tend to have a large $E_T^{miss}$.

\begin{figure}
\vspace{-2.0cm}
 \epsfig{file=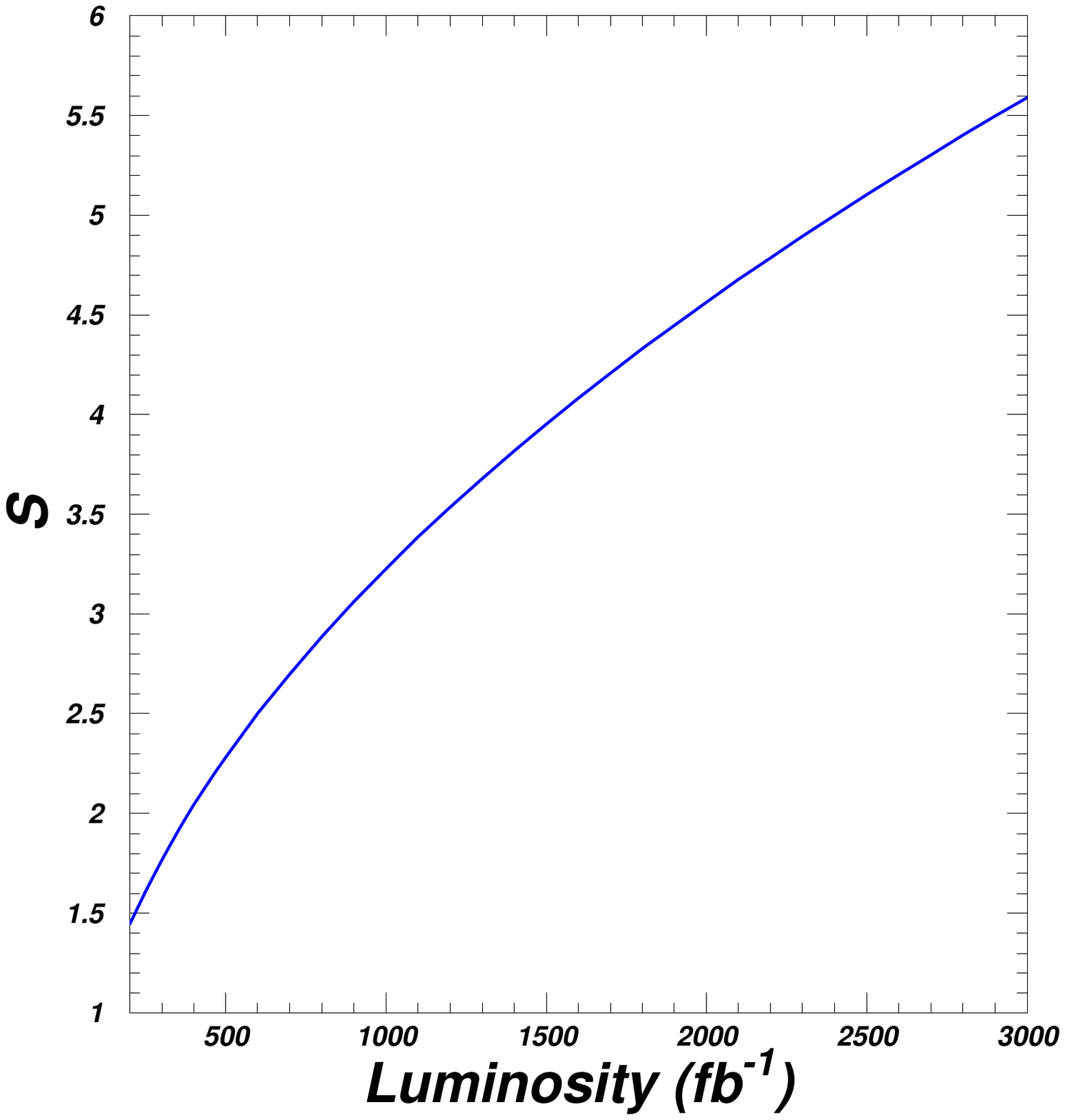,height=13.cm}
\vspace{-2.7cm} \caption{The significance versus the integrated luminosity of the 14 TeV LHC for the benchmark point.} \label{figlum}
\end{figure}

We compute the significance as $\mathbf{S}=\frac{n_s}{\sqrt{n_s+n_b}}$, where $n_s$ and $n_b$ are the normalized signal and background event yields, respectively. 
After making the kinematical cuts of Eq. (\ref{basiccut}), $n_b$ is drastically reduced, and dominated over by $n_s$.
For example, $n_s\sim 33$ and $n_s + n_b \sim 35$ for a dataset 3000 fb$^{-1}$ at the 14 TeV LHC. 
Fig. \ref{figlum} shows that for the benchmark point, the significance can reach 2$\sigma$ and 5.6$\sigma$ at the 14 TeV LHC
with an integrated luminosity  400 fb$^{-1}$ and 3000 fb$^{-1}$.

\section{Conclusion}
In this paper we study the capability of LHC to test DM, $Z^{\prime}$, and vector-like quark in a local 
$U(1)_{L_\mu-L_\tau}$ model in light of the $b\to s\mu^+\mu^-$ anomaly and the DM observables. We take $m_Q<$ 2 TeV and $m_{X_R}<$ 2 TeV,
and find that the $b\to s\mu^+\mu^-$ anomaly and the DM observables favor $m_{X_I}< 350$ GeV and $m_{Z^{\prime}}< 450$ GeV after imposing 
relevant constraints from theory and $b\to s$ flavor observables.
The current searches for jets and missing transverse momentum 
at the 13 TeV LHC with 139 fb$^{-1}$ integrated luminosity data exclude $m_Q<$ 1.7 TeV. 
Finally, we propose a novel channel of probing these new particles at the high luminosity LHC via 
the QCD process $pp \to D\bar{D}$ or $pp \to U\bar{U}$ followed by the
decay $D\to  s (b) Z'X_I$ or $U \to u (c) Z' X_I$ and then $Z'\to\mu^+\mu^-$. Taking a benchmark point of $m_Q$=1.93 TeV 
 $m_{Z^\prime}=170$ GeV, and $m_{X_I}=$ 145 GeV, we perform a detailed Monte Carlo simulation, and find that such benchmark point can be accessible at the 14 TeV LHC with an integrated luminosity 3000 fb$^{-1}$.

\section*{Acknowledgment}
We thank Biaofeng Hou for the helpful discussions. This work was supported by the National Natural Science Foundation
of China under grant 11975013, 11775025, and by the Natural Science Foundation of
Shandong province (ZR2017JL002 and ZR2017MA004).

\end{document}